\documentclass[12pt,preprint]{aastex}

\shorttitle{21-cm Emission from Minihalos}
\shortauthors{Iliev, Shapiro, Ferrara, Martel}
\begin{document}
\title{On the Direct Detectability of the Cosmic Dark Ages: 21-cm Emission
from Minihalos 
}
\author{Ilian T. Iliev\altaffilmark{1}, Paul R. Shapiro\altaffilmark{2},
 Andrea Ferrara\altaffilmark{1}, and Hugo Martel\altaffilmark{2}}
\altaffiltext{1}{
Osservatorio Astrofisico di Arcetri, Largo Enrico Fermi 5,
50125 Firenze, Italy}
\altaffiltext{2}{Department of Astronomy, University of Texas, Austin, TX 78712}
\email{iliev@arcetri.astro.it, shapiro@astro.as.utexas.edu}
\email{ferrara@arcetri.astro.it, hugo@simplicio.as.utexas.edu}

\begin{abstract} In
the standard Cold Dark Matter (CDM) theory of structure formation,
virialized minihalos (with $T_{\rm vir}\lesssim10^4$ K) form in abundance
at high redshift ($z>6$), during the cosmic ``dark ages.'' 
The hydrogen in these minihalos, the first nonlinear baryonic structures to
form in the universe, is mostly neutral and
sufficiently hot and dense to emit strongly at the 21-cm line. We calculate 
the emission from individual minihalos and the radiation background 
contributed by their combined effect. Minihalos create a ``21-cm forest''
of emission lines. We predict that the angular fluctuations in this
21-cm background should be detectable with the planned LOFAR 
and SKA radio arrays, thus providing a direct 
probe of structure formation during the ``dark ages.''
Such a detection will serve to confirm the basic CDM paradigm while
constraining the shape of the
power-spectrum of primordial density fluctuations down to much smaller
scales than have previously been constrained, 
the onset and
duration of the reionization epoch, and the conditions which led to the
first stars and quasars. We present results here for the currently-favored, 
flat $\Lambda$CDM model, for different tilts of the primordial power spectrum.
\end{abstract}


\section{Introduction}
No direct observation of the universe during the period between the 
recombination epoch at redshift $z\simeq10^3$ and the reionization 
epoch at $z\gtrsim6$ has yet been reported. While a number of suggestions
for the future detection of the reionization epoch, itself, have been made,
this period prior to the formation of the first stars and quasars -- the
cosmic ``dark ages'' (e.g. Rees 1999)
-- has been more elusive. Standard Big Bang cosmology 
in the CDM model predicts that nonlinear baryonic 
structure first emerges during this period, with virialized halos 
of dark and baryonic matter which span a range of masses from less than
$10^4M_\odot$ to about $10^{8}M_\odot$ which are filled with neutral hydrogen atoms. 
The atomic
density $n_H$ and kinetic temperature $T_K$ of this gas are high 
enough that collisions populate the hyperfine levels of the ground state of 
these atoms in a ratio close to that of their statistical weights (3:1), 
with a spin temperature $T_S$ that greatly exceeds the
excitation temperature $T_*=0.0681{\rm K}$. Since, as we shall show, 
$T_S>T_{\rm CMB}$, the temperature
of the Cosmic Microwave Background (CMB), as well,
for the majority of the halos,
these ``minihalos'' can be a detectable source of redshifted 21-cm line
emission. The direct detection of minihalos at such high redshift would be
an unprecedented measure of the density fluctuations in the baryons and
of the total matter power spectrum at small scales,
which will not be probed by other methods yet discussed 
(e.g. CMB anisotropy).

The possibility of 21-cm line emission or absorption by neutral H at
high redshift has been considered before 
\citep{HR79,SR90,SP93,KPS95,BNP97,MMR97,SWMB99,TMMR00}. 
Prior to the release of radiation by nonlinear baryonic 
structures which condense out of the background universe, the spin temperature
of $\rm H\,I$ 
in the diffuse, uncollapsed gas in the intergalactic medium (IGM) is
coupled to the CMB, so that $T_S=T_{\rm CMB}$ and neither emission nor 
absorption in the 21-cm line is possible.  
Recently, attention has focused on the possibility that radiation from
early stars and quasars might decouple $T_S$ from $T_{\rm CMB}$ by
``Ly$\alpha$ pumping'' -- resonant scattering in the H Ly$\alpha$ transition
followed by decay of the upper state $n=2$ to the ground state $n=1$ into
one or the other of the hyperfine levels 
(Madau et al. 1997; Tozzi et al. 2000).
This mechanism, it has been suggested, will operate on $\rm H\,I$ in
the diffuse, uncollapsed IGM during reionization, first to make
$T_S<T_{\rm CMB}$, so that the 21-cm transition can be seen in
{\it absorption} against the CMB, until the same Ly$\alpha$ scattering heats
the gas shortly thereafter and makes $T_S>T_{\rm CMB}$, thereby causing
21-cm {\it emission} in excess of the CMB, before reionization finally
destroys the $\rm H\,I$. In what follows, however, we show that a
substantial fraction of the baryons in the universe may already have 
condensed out of the diffuse IGM into virialized minihalos, {\it prior} to
and during reionization. Under these conditions,
collisional excitation alone is sufficient
to decouple $T_S$ from $T_{\rm CMB}$ and cause 21-cm emission in
excess of the CMB, thereby providing a signature of the cosmic ``dark
ages'' and of their retreat during reionization.

\section{21-cm Emission from Individual Minihalos}

The 21-cm emission from a single halo depends upon its internal 
atomic density,
temperature, and velocity structure. We model each CDM minihalo here as a
nonsingular, truncated isothermal sphere (``TIS'')
of dark matter and baryons in virial and hydrostatic equilibrium,
in good agreement with the results of gas and N-body simulations from
realistic initial conditions \citep{SIR99,IS01,IS02}.
This model uniquely specifies the internal structure of each halo (e.g. 
total and core sizes, $r_t$ and $r_0$, central total mass density $\rho_0$,
dark matter velocity dispersion $\sigma_V=(4\pi\rho_0r_0^2)^{1/2}$,
and gas temperature $T_K=\mu m_p\sigma_V/k_B$, where $\mu$ is
the mean molecular weight),
for a given background cosmology as functions of two parameters,
the total mass $M$ and collapse redshift $z_{\rm coll}$.

The minihalos which contribute significantly to the 21-cm emission
span a mass range from $M_{\rm min}$ to $M_{\rm max}$ which 
varies with redshift. $M_{\rm min}$ is close to 
the Jeans mass of the uncollapsed IGM prior to reionization,
$M_J=5.7\times10^3\left({\Omega_0h^2}/{0.15}\right)^{-1/2}
\left({\Omega_bh^2}/{0.02}\right)^{-3/5}\break\times
\left[{(1+z)}/{10}\right]^{3/2}M_\odot$, 
while 
$M_{\rm max}=3.95\times10^7\break\times
(\Omega_0h^2/0.15)^{-1/2}[(1+z)/10]^{-3/2}$
is the mass for which $T_{\rm vir}=10^4$ K according to the TIS model 
(Iliev \& Shapiro 2001)
(since halos with $T_{\rm vir}\gtrsim10^4{\rm K}$ are largely 
collisionally ionized).
Halos with $T_{\rm vir}\gtrsim10^4{\rm K}$ may have radiatively cooled 
gas inside them which would add to the signal we compute, but such gas is
expected to lead 
to the formation of internal sources of ionizing radiation which
will largely offset the effect. Since these additional effects are 
highly uncertain and are related to the onset of radiative feedback
and reionization which we are neglecting in these calculations, we 
will not consider the role of higher temperature halos further.

The flux per unit frequency,
${\mathfrak{F}}_\nu\equiv(dF/d\nu)_{\rm rec}$, received
at redshift $z=0$ at frequency $\nu_{\rm rec}$ from
a minihalo at redshift $z$ which emits at frequency 
$\nu_{\rm em}=\nu_{\rm rec}(1+z)$ is expressed in terms of the brightness
temperature $T_{b,\rm em}=T_{b,\rm rec}(1+z)$ according to
\begin{equation}
{\mathfrak{F}}_{\nu_{\rm rec}}
=\frac{2\nu_{\rm rec}^2}{c^2} k_B T_{b,\rm rec}(\Delta\Omega)_{\rm halo}\,,
\end{equation}
where $(\Delta\Omega)_{\rm halo}=\pi r_t^2/D_A^2
=\pi(\Delta\theta_{\rm halo}/2)^2$ is the solid
angle subtended by the minihalo, and $D_A$ is the angular diameter distance.
The brightness temperature $T_{b,\rm em}$
is determined by solving the equation of radiative
transfer to derive the brightness profile of the minihalo and
integrating this profile over the projected surface area, as follows.

The brightness temperature along a line of sight thru a minihalo at 
projected distance $r$ from the center obeys the equation
\begin{equation}
\label{T_b_full}
T_b(r)=T_{\rm CMB}e^{-\tau(r)}
	+\int^{\tau(r)}_{0}T_S e^{-\tau'}d\tau'\,,
\end{equation}
where quantities are defined in the comoving frame of
the minihalo,
frequency
$\nu$ refers here and henceforth to $\nu_{\rm em}$, $\tau(r)$ is the 
total optical depth thru the halo, and
the effective absorption coefficient $\kappa_\nu$ is given 
when $T_*\ll T_S$ by:
\begin{equation}
\label{kappa}
\kappa_\nu = \frac{3c^2A_{10}n_{\rm HI}}{32\pi\nu^2} f(\nu)\frac{T_*}{T_S}
\end{equation}
\citep{F58},
where $A_{10}=2.85\times10^{-15}{\rm s}^{-1}$ 
is the Einstein $A$-coefficient for the 21-cm transition and
$f(\nu)$ is the normalized line profile. 
The spin temperature $T_S$ is determined by the balance between
collisional and radiative excitation and de-excitation
by atoms and electrons and by CMB and Ly$\alpha$ photons, respectively,
according to
\begin{equation}
\label{T_S_2}
T_S = \frac{T_{CMB}+y_cT_K+y_\alpha T_\alpha}{1+y_c+y_\alpha}\,,
\end{equation}
where $T_\alpha$ is the color temperature of the Ly$\alpha$
photons, and $y_\alpha$ and $y_c$ are radiative and collisional excitation
efficiencies, respectively \citep{PF56,F58,F59}. 
The efficiency $y_c$ includes contributions from
$\rm H^0-H^0$ collisions, $y_H$, and from $\rm e^--H^0$ collisions,
$y_e$.
Prior to the reionization epoch, Ly$\alpha$ pumping is unimportant and
collisional excitation alone must compete with excitation by the CMB. 
This is only possible for gas that is highly nonlinear
and sufficiently hot. Such conditions are achieved only inside
virialized halos.
As shown in Figure 1,
the optical depth of an individual halo is not 
negligible, particularly for smaller-mass
halos (due to their lower $T_S$).
Since $T_S$ varies with radial position inside the halo, as a result of its
significant central concentration, we must integrate equation~(\ref{T_b_full}) 
numerically.
The face-averaged $T_b$ of this single halo is given by
$\langle T_b\rangle_{\rm halo}\equiv(\int T_b(r)dA)/A$,
where $A(M,z)$ is the geometric cross-section of
a halo of mass $M$ and collapse redshift $z$.
The observed flux from an individual halo
is then expressed with respect to the CMB by the differential antenna temperature 
$\delta T_b\equiv[\langle T_b\rangle_{\rm halo}-T_{\rm CMB}(z)]/(1+z)$.

The line-integrated flux $F(M,z)$ received from this
minihalo is equal to the
flux calculated for $\nu=\nu_0$ multiplied by
a redshifted effective line-width $\Delta\nu_{\rm eff}(z)$, defined by
$\Delta\nu_{\rm eff}(z)
\equiv(\int{\mathfrak{F}}d\nu)/{\mathfrak{F}}_{\nu_0}$.
For an optically thin minihalo,
$\Delta\nu_{\rm eff}$ reduces to
$\Delta\nu_{\rm eff}(z)=[f(\nu_0)(1+z)]^{-1}$.
In that case, for a thermal-Doppler-broadened line profile,
$\Delta\nu_{\rm eff}(z) =[(2\pi\mu)^{1/2}\nu_0\sigma_V/c](1+z)^{-1}$.
We have checked that this approximation is adequate
even for the optically thicker halos at the
small-mass end of the mass function. 
The differential line-integrated flux $\delta F(M,z)$
is given by replacing $T_{b,\rm rec}$ in equation~(1) by $\delta T_b$
and integrating over frequency as described above.

Our results for individual minihalos are summarized
in Figure~\ref{6panel}.
Line profiles of different minihalos along the same line of sight should
not typically overlap. The proper mean free path 
$\lambda_{\rm mfp}=\langle n_{\rm halo}\sigma_{\rm halo}\rangle^{-1}$ for
photons to encounter minihalos in $\Lambda$CDM is 160~kpc at $z=9$
(Shapiro 2001), corresponding to a frequency separation, 
$\Delta\nu_{\rm sep}
\approx \nu_0 H(z)\lambda_{\rm mfp}/[c(1+z)]\sim0.1\,{\rm MHz}\gg
\Delta\nu_{\rm eff}\lesssim10\,{\rm kHz}$. 

These results predict a ``21-cm forest'' 
of minihalo emission lines.
At $z=9$, for example, there are about 160 minihalo lines per unit redshift
along a typical line of sight in an untilted $\Lambda$CDM universe
(Shapiro 2001). Detecting the stronger lines
would require sub-arcsecond spatial resolution, $\sim1\,\rm kHz$
frequency resolution, and $\sim\,\rm nJy$ sensitivity. SKA is expected to
have sufficient resolution for such observation, but probably not sufficient 
sensitivity.  
 
\section{21-cm Radiation Background from Minihalos}

The average differential flux per unit frequency
relative to that of the CMB
from all the minihalos observed within a given beam of angular size 
$\Delta\theta_{\rm beam}$ 
and frequency bin $\Delta\nu_{\rm obs}$ is:
\begin{equation}
\label{S_nu}
\overline{\delta{\mathfrak{F}}}_\nu(z)=
\frac{\Delta z(\Delta\Omega)_{\rm beam}}{\Delta\nu_{\rm obs}}
\frac{d^2V(z)}{dz\,d\Omega}
\int_{M_{\min}}^{M_{\max}}\delta F\frac{dn}{dM}dM,
\end{equation}
where ${d^2V(z)}/{dz\,d\Omega}$ is the comoving volume per unit 
redshift per unit 
solid angle, the solid angle 
$(\Delta\Omega)_{\rm beam}=\pi(\Delta\theta_{\rm beam}/2)^2$,
and $\Delta\nu_{\rm obs}/\Delta z=\nu_0/(1+z)^2$.
We calculate the comoving density of halos at different redshifts using the
Press-Schechter (PS) approximation for the halo mass function $dn/dM$. 
If we define the beam-averaged
``effective'' differential antenna temperature $\overline{\delta T}_b$ using
$\overline{\delta{\mathfrak{F}}}_\nu=2\nu^2k_B
\overline{\delta T_b}(\Delta\Omega)_{\rm beam}/c^2$, then 
\begin{equation}
\label{bar_T_b}
\overline{\delta T}_b=\frac{c(1+z)^4}{\nu_0 H(z)}
\int_{M_{\min}}^{M_{\max}}\Delta\nu_{\rm eff}\delta 
T_{b,\nu_0}A\frac{dn}{dM}dM.
\end{equation}
We consider the currently-favored,
flat CDM model with cosmological constant
(``$\Lambda$CDM'', $\Omega_0=0.3$, $\lambda_0=0.7$, 
COBE-normalized, $\Omega_bh^2=0.02$, $h=0.7$), for
three values of the primordial power spectrum 
index $n_p=0.9$, 1, and 1.1, using the 
primordial power spectrum of \cite{EH99}.

Results for $\overline{\delta{\mathfrak{F}}}_\nu$
and $\overline{\delta T}_b$ are plotted in Figure~\ref{tot_flux_z}.
In principle, the variation of $\overline{\delta T_b}$ with
observed frequency implied by the redshift variations in 
Figure~\ref{tot_flux_z} should permit a discrimination between
the 21-cm emission from minihalos and
the CMB and other backgrounds, due to their very different frequency
dependences.
However, the average differential brightness temperature of this
minihalo background is   
very low and its evolution is fairly smooth, so such measurement may be 
difficult in practice with currently planned instruments like LOFAR and SKA. 
The angular fluctuations in this emission, on the other hand,
should be much easier to detect, as discussed in the next section.

\section{Angular Fluctuations in the 21-cm Emission Background}

The amplitude of $q$-$\sigma$ angular fluctuations 
(i.e. $q$ times the rms value) in the differential 
antenna temperature is given in the linear regime by
\begin{equation}
\frac{\langle\delta T_b^2\rangle^{1/2}}{\overline{\delta T_b}}
	= qb(z)\sigma_p,
\end{equation}
where $\sigma_p$ is the rms mass fluctuation at redshift $z$ 
in a randomly placed cylinder which corresponds to
the observational volume defined by the
detector angular beam size, $\Delta\theta_{\rm beam}$, and 
frequency bandwidth, $\Delta\nu_{\rm obs}$, and $b(z)$ is 
the bias factor which accounts for the clustering of rare density peaks
relative to the mass.
We assume $b(z)$ is the flux-weighted average over the mass function of
$b(M,z)=1+(\nu_{\rm h}^2-1)/\delta_c$, the linear bias factor,
where $\nu_{\rm h} = \delta_c/\sigma(M)$, $\delta_c$
is the value of the linearly extrapolated value
of overdensity $\delta\rho/\rho$ corresponding to the epoch when a top-hat 
collapse reaches infinite density, and $\sigma(M)$ is the the standard 
deviation of the density contrast filtered on mass scale $M$ 
(e.g. Mo \& White 1996). For a cylinder of comoving 
radius $R=\Delta\theta_{\rm beam}(1+z)D_A(z)/2$, and length 
$L\approx(1+z)cH(z)^{-1}(\Delta\nu/\nu)_{\rm obs}$,
we have :
\begin{eqnarray}
\sigma_p^2&=&\frac{8D^{-2}(z)}{\pi^2R^2L^2}\int^\infty_0\!dk
\int_0^1\!dx
\frac{\sin^2(kLx/2)J_1^2[kR(1-x^2)^{1/2}]}{x^2(1-x^2)}
\nonumber\\
&&\hskip100pt\times(1+fx^2)^2\frac{P(k)}{k^2}
\end{eqnarray}
\citep{TMMR00} (with several typos in the corresponding
expression in that paper corrected here), where $D(z)\equiv\delta_+(0)/\delta_+(z)$ 
is the linear growth factor, $P(k)$ is the linear power spectrum at $z=0$,
and the factor $(1+fx^2)^2$, with
$f\approx[\Omega(z)]^{0.6}$,
is the correction to the cylinder length for the departure from Hubble
expansion due to peculiar velocities \citep{K87}.

Illustrative results are plotted 
for 3-$\sigma$ fluctuations
as a function of $\Delta\theta_{\rm beam}$, for $z=7$ and 8.5,
in Figure~\ref{rms_z8.5}, along with
the expected sensitivity limits for the planned LOFAR (300 m filled 
aperture) and SKA (1 km filled aperture) arrays.
We plot in Figure~\ref{rms_theta25} the predicted spectral variation of these
fluctuations vs. redshift $z$ for illustrative
beam sizes of $\Delta\theta_{\rm beam}=9'$ and $25'$.
These 3-$\sigma$ fluctuations should be observable with both LOFAR 
and SKA with
integration times of between 100 and 1000 hours. 
For a $25'$ beam, for example,
3-$\sigma$ fluctuations can be detected for untilted $\Lambda$CDM by both
with a 100~h integration for $z\sim6-7.5$ and 
a 1000~h integration for $z\lesssim11.5$, while for a $9'$ beam,
SKA can detect them after 100~h for $z\lesssim9$ and
after 1000~h for $z\lesssim13$. Results for different values
of $z$ and $\Delta\theta_{\rm beam}$ are available upon request.  

\section*{Acknowledgments}
We are grateful to E. Scannapieco for clarifying the effects of bias
and referee P. Tozzi for his thoughtful comments.
This work was supported by European
       Community RTN contract HPRN-CT2000-00126 RG29185
and grants NASA ATP NAG5-10825 and NAG5-10826 and Texas
Advanced Research Program 3658-0624-1999.

\figcaption[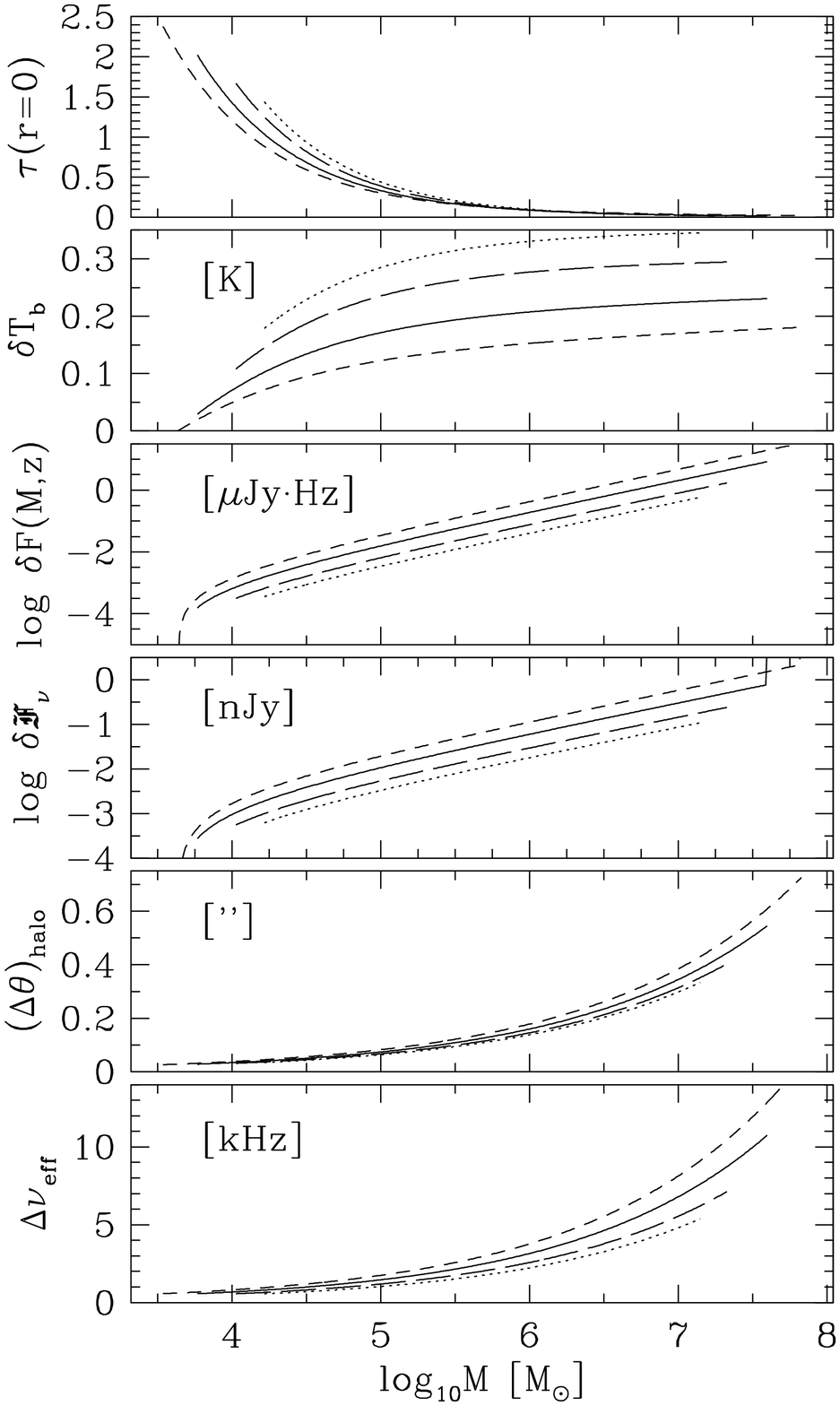]
{Individual minihalo sources of redshifted 21-cm emission
in $\Lambda$CDM,
redshifts $1+z=7$ (short-dashed line), 10 (solid line), 15 (long-dashed line),
and 20 (dotted line) vs. total mass of minihalo $M$.
From top to bottom: optical depth $\tau_{\nu_0}(r=0)$ at 
line-centered frequency $\nu_0$ thru minihalo center,
differential antenna temperature $\delta T_b$,
line-integrated differential flux $\delta F(M,z)$ relative to the CMB,
total differential flux  per unit frequency ${\mathfrak{F}}_{\nu_0}$,
angular size of minihalo $(\Delta\theta)_{\rm halo}$,
and redshifted effective width $\Delta\nu_{\rm eff}(z)$
of the 21-cm line as observed at $z=0$ at received frequency
$\nu_{\rm rec}=\nu_0(1+z)^{-1}$.
\label{6panel}}

\figcaption[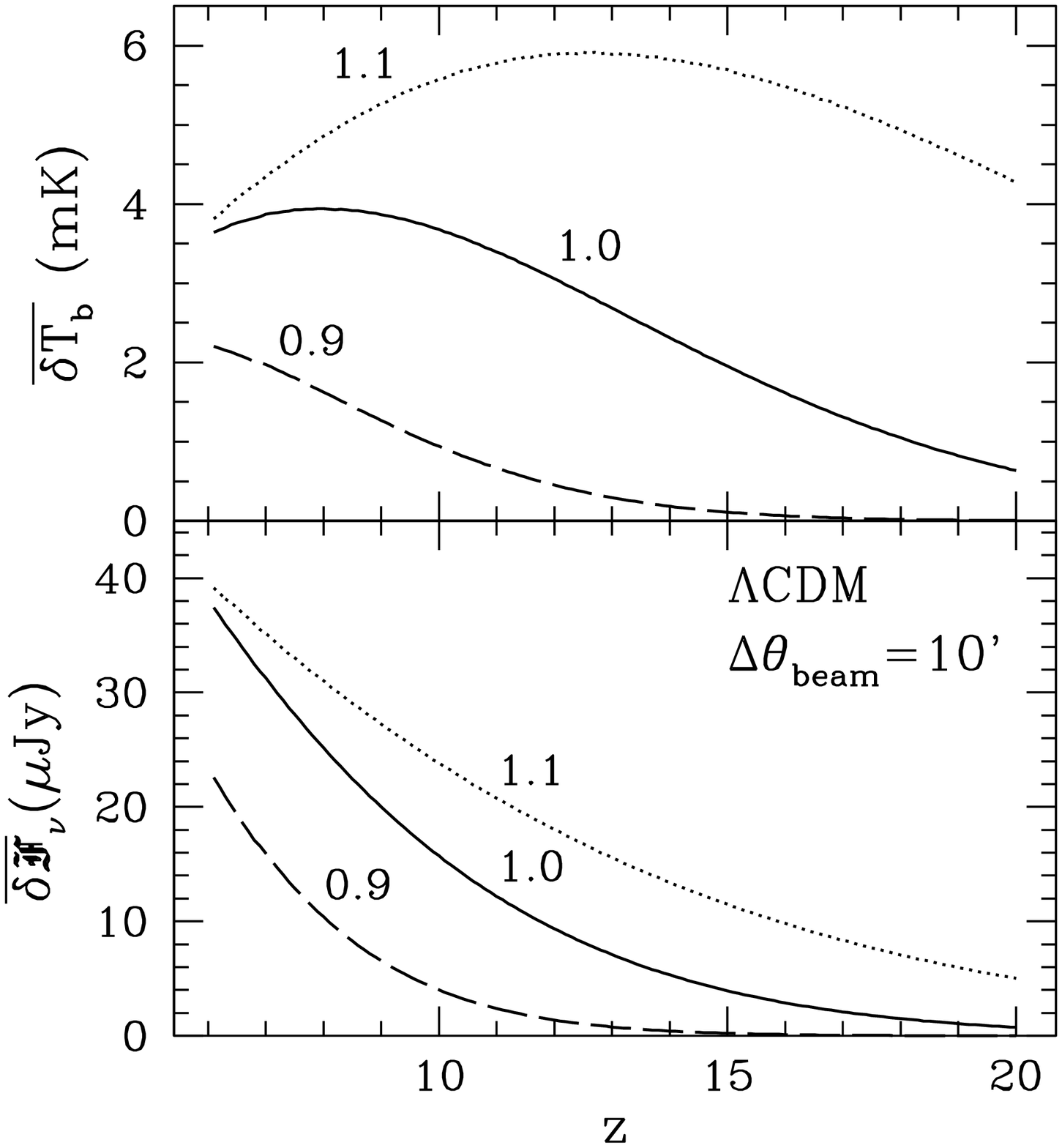]
{Minihalo radiation background.
Average observed differential antenna temperature $\overline{\delta T}_b$
and average differential flux per unit frequency 
$\overline{{\delta\mathfrak{F}}}_\nu$ for beam size of $\Delta\theta_{\rm beam}=10'$ 
at the redshifted 21-cm line frequency due to minihalos vs. redshift $z$ 
for $\Lambda$CDM models with power-spectrum tilts
$n_p=0.9$, 1.0, and 1.1, as labelled.
\label{tot_flux_z}}

\figcaption[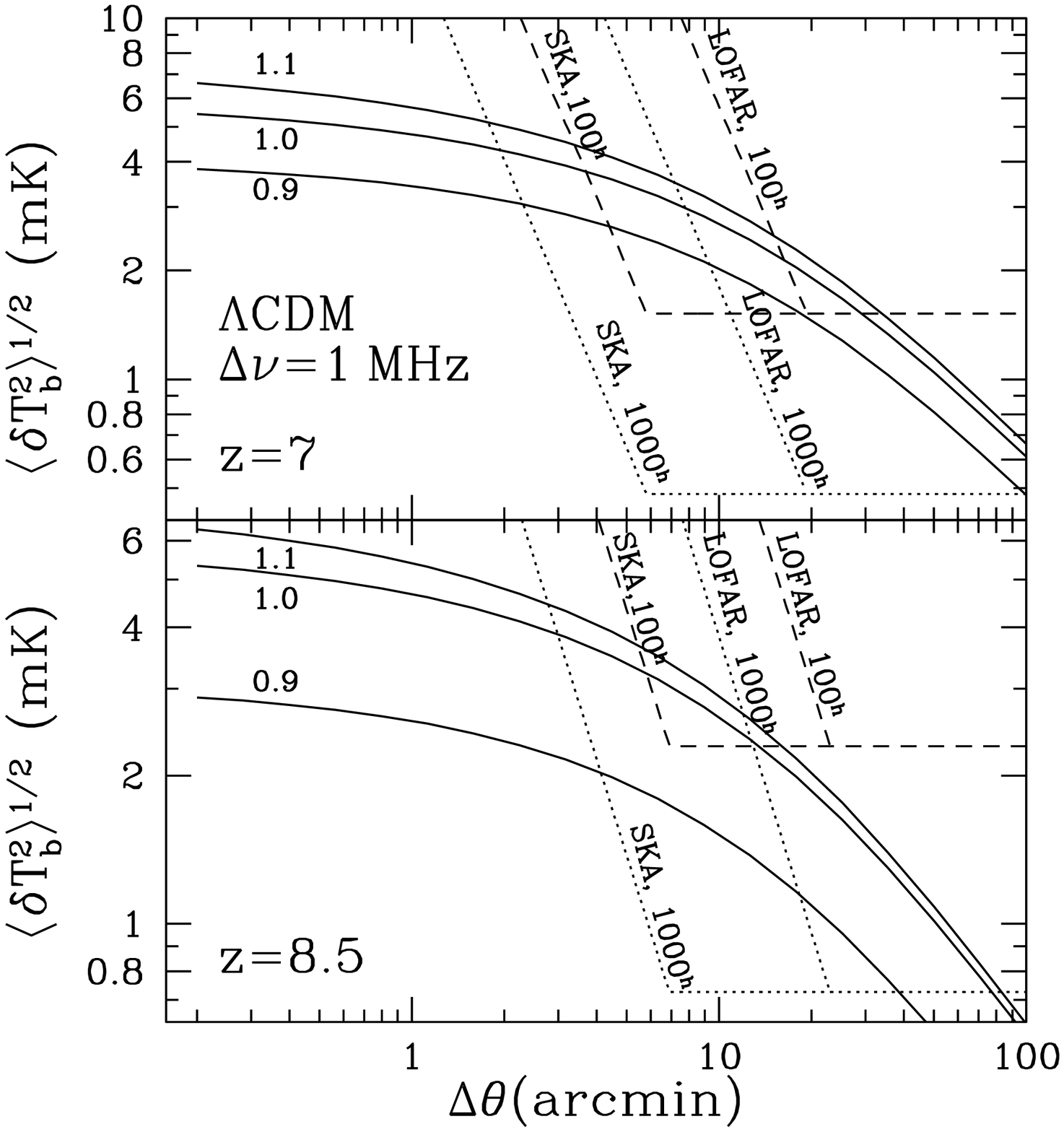]
{Predicted 3-$\sigma$
differential antenna temperature fluctuations
at $z=7$ (($\nu_{\rm rec}=177.5$ MHz; top panel) 
and $z=8.5$ ($\nu_{\rm rec}=150$ MHz; bottom 
panel) for bandwidth $\Delta\nu_{\rm obs}=1\,\rm MHz$ 
vs. angular scale $\Delta\theta_{\rm beam}$ 
for $\Lambda$CDM models with tilt 
$n_p=0.9$, 1.0, and 1.1, as labelled (solid curves).
Also indicated is the predicted sensitivity of LOFAR and SKA 
for a confidence level of 5 times the noise level after
integration times of
100~h (dashed lines) and 1000~h (dotted lines), 
with compact subaperture (horizontal lines) and extended
configuration needed to achieve higher resolution (diagonal lines) 
(see http://www.lofar.org/science).
\label{rms_z8.5}}

\figcaption[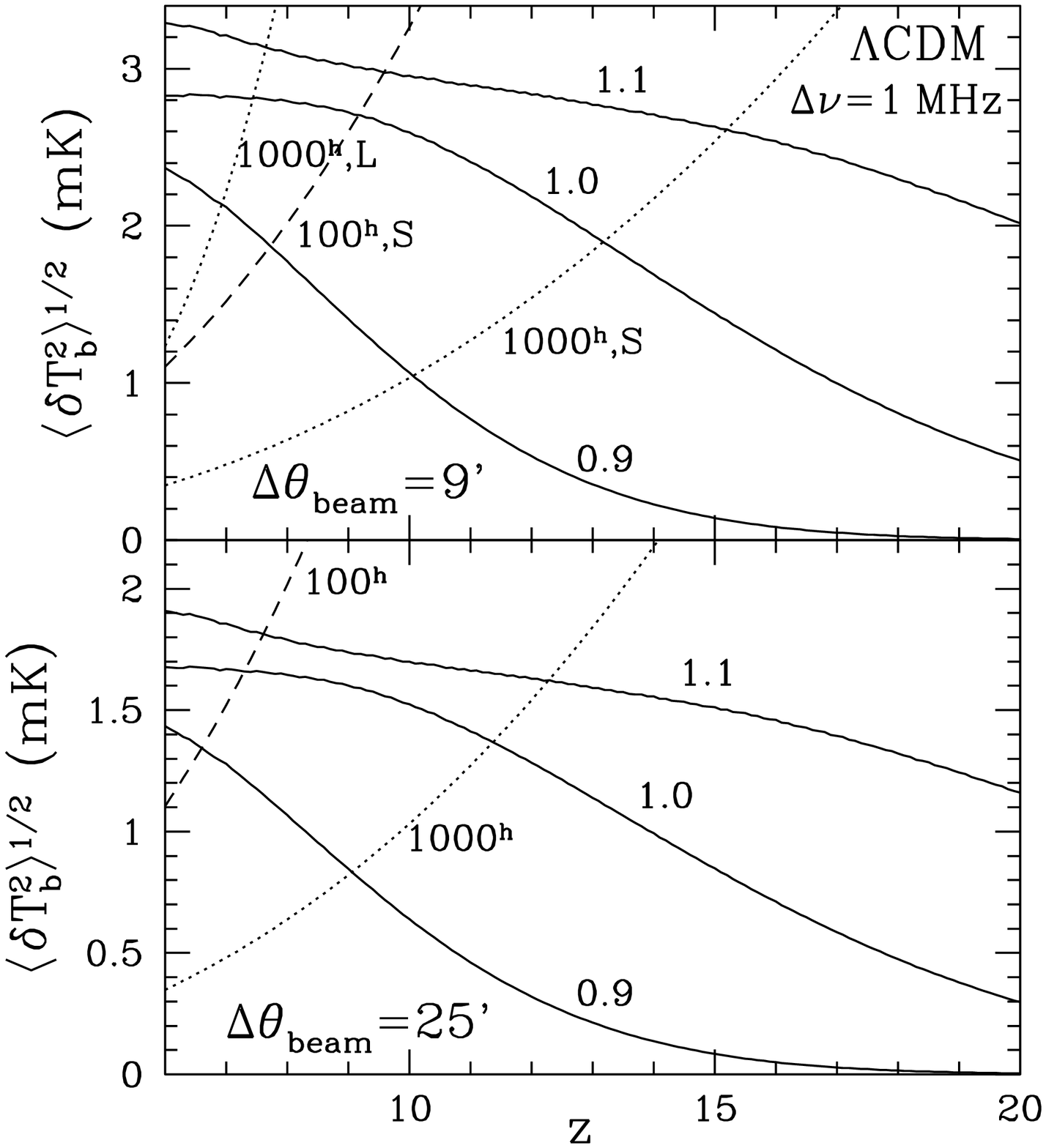]{
Predicted 3-$\sigma$ 
differential antenna temperature fluctuations
at $\Delta\theta_{\rm beam}=9'$ (top panel) and $25'$ (bottom panel) vs.
 redshift $z$ for $\Lambda$CDM models with tilt 
$n_p=0.9$, 1.0, and 1.1, as labelled (solid curves).
As in Figure~3, we also plot
the predicted sensitivity  for integration times 100~h 
(dashed) and 1000~h (dotted)
of both LOFAR (``L'') and SKA (``S''), as labelled
(for bottom panel, sensitivity curves for LOFAR and SKA are identical),
for compact subaperture and assuming rms sensitivity 
$\propto \nu^{-2.4}$ (see http://www.lofar.org/science).
\label{rms_theta25}}

\begin{figure}
\epsscale{0.5}
\plotone{f1.eps}
\end{figure}

\begin{figure}
\epsscale{0.6}
\plotone{f2.eps}
\end{figure}

\begin{figure}
\plotone{f3.eps}
\end{figure}

\begin{figure}
\plotone{f4.eps}
\end{figure}
\end{document}